\documentclass[showkeys,eqsecnum,showpacs,prd,superscriptaddress]{revtex4}
\usepackage[dvips]{graphicx}
\usepackage{color}
\usepackage{amsmath}
\usepackage[dvips]{graphicx}
\begin{document}
\newcommand{\be}{\begin{equation}}
\newcommand{\ee}{\end{equation}}

\title{Gravitation and Thermodynamics: The Einstein Equation of State Revisited}

\author{Jarmo M\"akel\"a} 
\email[Electronic address: ]{jarmo.makela@puv.fi}  
\affiliation{Vaasa University of Applied Sciences, Wolffintie 30, 65200 Vaasa, Finland}
\author{Ari Peltola}
\email[Electronic address: ]{a.peltola-ra@uwinnipeg.ca} 
\affiliation{Department of Physics, The University of Winnipeg, 515 Portage Avenue, Winnipeg, Manitoba. Canada. R3B 2E9}

\begin{abstract} 
We perform an analysis where Einstein's field equation 
is derived by means of very simple thermodynamical arguments. 
Our derivation is based on a consideration of the properties of a very
small, spacelike two-plane in a uniformly accelerating 
motion.
  
\end{abstract}


\maketitle

\section{Introduction}
Ever since the discovery of the Bekenstein-Hawking entropy law, it has
become increasingly clear that there is a deep connection between
gravitation and thermodynamics (see, for instance, Refs. 
\cite{bek,haw,unruh,bath1,pad}). However, even today it is not
properly understood what exactly this connection may be. The most
surprising point of view on these matters was probably provided by
Jacobson in 1995, when he discovered that Einstein's field equation is
actually a thermodynamical equation of state of spacetime and matter
fields \cite{jac}. The key point in his analysis was to require that
the first law of thermodynamics, which implies the fundamental
thermodynamical relation
\be \label{eq:sinikka1} \delta Q = T\, dS,
\ee
holds for all local Rindler horizons, and that the entropy $S$ of a
finite part of the Rindler horizon is one-quarter of its area. 
Jacobson considered an observer very close to his local Rindler 
horizon (which means that the proper acceleration $a$ of the observer is extremely 
large). For the temperature $T$ in Eq. (\ref{eq:sinikka1}), Jacobson took the Unruh 
temperature
\be \label{eq:unruhT} T_\text{U} = \frac{a}{2\pi}
\ee
experienced by the observer, and the heat flow $\delta Q$ through the
past Rindler horizon was
defined to be the boost-energy current carried by matter. Jacobson was 
able to show that, under the assumptions mentioned above, the heat flow 
through the horizon causes a decrease in the horizon area in such a way 
that Einstein's field equation is satisfied. In other words, he was able to
derive Einstein's field equation by assuming the first law of 
thermodynamics and the proportionality of entropy to the 
area of the horizon. Viewed in this way, Einstein's field equation is 
nothing more than a thermodynamical equation of a state \cite{edd,tiw}.

The purpose of this paper is to investigate whether there are some other
(possibly more general) principles of nature that would imply
Einstein's field equation. Recently, it has been suggested that the
concept of gravitational entropy should be extended from horizons to
arbitrary spacelike two-surfaces with finite areas \cite{mp1,mp2,jamo}. 
In Ref. \cite{jamo} it was proposed that an accelerated two-plane may
be associated with an entropy which is, in natural units, one-half of
the area of that plane. This proposal is, in some sense, 
related to the well-known result that the entropy associated with a
spacetime horizon is one-quarter of the area of the horizon. The
reason for the difference in the constant of proportionality is still
unclear, but it may result from the fact that a spacetime horizon is,
according to observers having that surface as a horizon, only a
one-sided surface, whereas an accelerated spacelike two-surface has two
sides \cite{thie}.

In this paper we shall find that Einstein's field equation can be
derived from a hypothesis which is closely related to this proposal. 
Our derivation will be based on a consideration of a very small, 
spacelike two-plane accelerating uniformly in a direction perpendicular 
to the plane. When the plane moves in spacetime with respect to the matter 
fields, matter will flow through the plane. Since the matter has, from
the point of view of an observer at rest with respect to the plane, a certain
non-zero temperature, it also has a certain entropy content. In other
words, entropy flows through the plane. Since the plane is in an accelerating
motion, the entropy flow through the plane (amount of entropy flown through the 
plane in unit time) is not constant, but it will change as a function of the proper 
time of an observer moving along with the plane.

The change in the entropy flow through the plane has two parts. One of these parts
is due to the simple fact that the plane moves from one point to another in spacetime,
and the entropy densities in the different points of spacetime may be different. This
part has nothing to do with the acceleration of the plane. Another part in the change
of the entropy flow, however, is caused by the change in the velocity of the plane 
with respect to the matter fields: When the velocity of the plane with respect to the 
matter fields changes, so does the entropy flow through the plane. This part in the 
change of the entropy flow is caused by the acceleration of the plane, and it is this
 part in the change of the entropy flow, where we shall focus our attention. For the sake of
brevity and simplicity we shall call that part as the change in the \emph{acceleration
entropy flow}.

When the accelerating plane moves in curved spacetime, its area may
change. More precisely, when the accelerating plane moves in curved 
spacetime, the world lines of the points of the plane may either approach
to each other or recede from each other. To investigate the behaviour of those
world lines, we shall consider a congruence of timelike curves with certain
specific properties. The physical idea behind our consideration is that when
our plane is located at a certain spacetime point $\mathcal{P}$, then in the 
local neighbourhood of that point the world lines of the points of our 
accelerating plane are the elements of that congruence. In 
broad terms, we shall define this congruence in such a way that in the immediate 
vicinity of the point $\mathcal{P}$ the tangent vectors of the world lines of 
the points of our plane are parallel to each other, and all points of all elements 
of the congruence are accelerated with a constant proper acceleration in a
direction perpendicular to the plane. Such a definition is consistent 
with the intuitive picture of the concept of a two-plane moving in spacetime. 
These ideas will become more precise in the Section \ref{sec:area} of this 
paper, when we consider the change of the area of our plane. 
Because the tangent vectors of the world lines of the points of the 
plane are defined to be parallel to each other in the instant vicinity 
of the point $\mathcal{P}$, the proper time derivative of the area $A$ of the 
plane will vanish at that point. In other words, if we parametrize the world 
lines of the points of our plane by means of the proper time $\tau$ measured 
along those world lines such that $\tau=0$, when the plane lies at the point 
$\mathcal{P}$, we must have $\frac{dA}{d\tau}\vert_{\tau=0}=0$. If spacetime 
is curved, however, $\frac{dA}{d\tau}$ will become non-zero, when $\tau>0$. 
From this point on, we shall call the quantity $-\frac{dA}{d\tau}$ as the 
\emph{shrinking speed} of the area of the plane.

Under the assumption that the rate of change in the boost energy flow through the 
plane is exactly 
the the rate of change in the heat flow, we express the following hypothesis concerning the rates of changes 
in the acceleration entropy flow through an accelerating plane, and in the shrinking 
speed of the area of the plane:

\emph{If the temperature of the matter flowing through an accelerating, spacelike
two-plane is equal to the Unruh temperature measured by an observer at rest with 
respect to the plane, then the rate of change in the acceleration entropy flow through 
the plane is, in natural units, exactly one-half of the rate of change in the
shrinking speed of the area of the plane.} 

Using this hypothesis, and this hypothesis only, together with
Eq. (\ref{eq:sinikka1}), we shall obtain Einstein's field
equation. Our hypothesis may be
expressed by means of a formula: 
\begin{equation} \label{eq:rissi}
\frac{d^2S_a}{d\tau^2}\Big|_{\tau=0} = -\frac{1}{2}\frac{d^2A}{d\tau^2}\Big|_{\tau=0},
\end{equation}
where $\frac{dS_a}{d\tau}$ denotes the acceleration entropy flow, and 
$-\frac{dA}{d\tau}$ the shrinking speed of the area. Because the Unruh
temperature $T_\text{U}$ of Eq. (\ref{eq:unruhT}) represents, in some
sense, the temperature of spacetime from the point of view of an
observer moving with a constant proper acceleration $a$, we may view 
Eq. (\ref{eq:rissi}) as an equation which holds, when matter and
spacetime are, from the point of view of an accelerating observer, in
a thermal equilibrium with each other. When we calculate the rate of
change in the acceleration entropy flow through the plane, we must  
use Eq. (\ref{eq:sinikka1}). More precisely, we first calculate the
rate of change in the flow of heat through the plane and then, using
Eq. (\ref{eq:sinikka1}) and identifying $T$ as the Unruh temperature
$T_\text{U}$ of Eq. (\ref{eq:unruhT}), we calculate the rate of change
in the acceleration entropy flow. We have assumed that the 
rate of change in the boost energy
flow through our accelerating plane is exactly the rate of 
change in the heat flow for the
simple reason that it makes the calculation of the flow of entropy
very easy: We just calculate the rate of change in the boost energy flow, 
and then use
Eq. (\ref{eq:sinikka1}). If there were other forms of energy, except
heat, flowing through our plane, it would not be quite clear what we
actually mean by the concept of entropy flow, and our analysis would
become much more complicated. It is most gratifying that Einstein's
field equation follows from our hypothesis even with this rather
restrictive assumption, regardless of what kind of matter we happen to
have. 

It is important to note that our hypothesis contained in Eq. (\ref{eq:rissi})
involves \emph{second} proper time derivatives only. The reason for this is
easy to understand: If the initial velocity of an accelerated plane with
respect to the matter fields is undetermined, the acceleration of the plane
contributes to the \emph{rate of change} of the entropy flow through
the plane only, and not to the entropy flow itself, which depends on the velocity
of the plane only. Hence we must consider the second, instead of the first
proper time derivative of the amount of entropy carried through the plane. For
similar reasons we must consider the rate of change of the shrinking speed of
the plane, instead of the shrinking speed itself: In a given point of spacetime
the shrinking speed of the plane depends on the initial conditions given for
the tangent vectors of the world lines of the plane only, and therefore it has
no direct connection with the geometric properties of spacetime. In contrast, the
rate of change in the shrinking speed, or the negative of the second proper time 
derivative of the area, does indeed depend on the spacetime geometry. Hence we 
may conclude that if we want to find the relationship between acceleration, 
entropy and spacetime geometry, we must consider the second, instead of the 
first proper time derivatives of the area and entropy.

We begin our investigations in Sec. \ref{sec:tra} by considering the
trajectory of our plane. We shall assume that at a certain point
$\mathcal{P}$ of spacetime we have an orthonormal geodesic frame of
reference, where all components of the energy momentum stress tensor
$T_{\mu\nu}$ of matter are fixed and finite. In this frame of reference 
we shall then introduce a very small spacelike two-plane,
which moves, at the point $\mathcal{P}$, with a velocity very close to
the speed of light to the direction of its normal and, at the same
time, accelerates with a constant proper acceleration to the opposite
direction. In order to make our analysis sufficiently local, the
proper acceleration of the plane is taken to be very large. For
sufficiently large values of the proper acceleration, one may view the
local neighbourhood of the point $\mathcal{P}$ as a region of
spacetime which possesses the ordinary properties of the Rindler
spacetime, including the Unruh temperature $T_\text{U}$ of
Eq. (\ref{eq:unruhT}). 

In Sec. \ref{sec:area} we shall focus our attention to the change in
the area of our accelerating two-plane. As the final
result of Sec. \ref{sec:area} we shall get the rate of change in the
shrinking speed of the plane. 

The motivation for our decision to consider a plane moving with a very
high speed in a chosen frame of reference becomes obvious in
Sec. \ref{sec:flow}, where we consider the flow of heat through our
accelerating plane. It is fairly easy to show that if matter consists
of a gas of non-interacting massless particles, i.e., of massless  
radiation, then the flow of boost energy through the plane is exactly
the heat flow through the plane. Unfortunately, if the particles
of the matter fields are massive, the situation becomes more
complicated, because in that case other forms of energy, except heat,
(mass-energy, for instance) are carried through the plane. However, if
the plane moves with an enormous velocity with respect to the matter
fields, then the kinetic energies of the particles of the fields
vastly exceed, in the rest frame of the plane, all the other forms of
energy. In this limit we may consider matter, in effect, as a gas of  
non-interacting massless particles, and the rate of change in the
boost energy flow is exactly the rate of change in the heat flow. 
We identify that part in the rate of change in the
heat flow, which is due to the mere acceleration of the plane, and using
Eq. (\ref{eq:sinikka1}) we calculate the rate of change in the
acceleration entropy flow. 

After obtaining an expression for the rate of change in the
acceleration entropy flow in Sec. \ref{sec:flow}, and for the rate of
change in the shrinking speed in Sec. \ref{sec:area}, we are finally
able to obtain, in Sec. \ref{sec:rad}, Einstein's field equation by
means of our hypothesis in the special case, where matter consists of 
massless, non-interacting radiation fields (electromagnetic field, for example),
which are initially in a thermal equilibrium in the rest frame of our plane.
In Sec. \ref{sec:EFE} it is found that Einstein's field equation for general 
matter fields is a straightforward consequence of our hypothesis in the limit, where the
plane moves with a velocity very close to the speed of light with
respect to the matter fields .

We close our discussion in Sec. \ref{sec:koonti} with some concluding
remarks. 

\section{Trajectory of the Plane} \label{sec:tra}
It is now time to specify our thermodynamical system in detail. 
Take a spacetime point $\mathcal{P}$ and define an orthonormal
geodesic system of coordinates $t,x,y,z$ at the local neighbourhood of
that point. The origin of the coordinates is taken to lie
at $\mathcal{P}$. Consider then a uniformly accelerated observer with
a proper acceleration 
$a$ travelling through $\mathcal{P}$ in the direction of the positive
$z$-axis. We denote the velocity of that observer at $\mathcal{P}$ by
$v>0$. Furthermore, we assume that the acceleration of that observer
is directed (in space) towards the negative $z$-axis. With the 
accelerating observer we shall now associate a small accelerated
two-plane in the following way: In the local neighbourhood surrounding
the observer, it is possible to define the concept of a two-plane. We
consider a small two-plane which always remains at rest with
respect to the observer. This means that at every point of the world
line of the observer, we visualize a certain spacelike two-plane,
constantly moving along with the observer. We assume that this
two-plane is perpendicular to the $z$-axis, which means that the
acceleration is directed perpendicular to the plane. 

There are obvious physical
reasons to require that the proper acceleration of the plane must be
very large. When spacetime is curved, one may associate
the ordinary Rindler wedge of the accelerating observer with the local
neighbourhood of the point $\mathcal{P}$ only. Hence, if we want to
employ the properties of Rindler spacetime in our calculations, we
must analyze the thermodynamics of the plane in the limit where the
proper acceleration $a$ becomes very large. However, we shall not
specify the actual magnitude of the proper acceleration in more
detail.  When the curvature of spacetime is reasonable large, one may
always make the analysis sufficiently local by increasing the value of
$a$. Only in very special circumstances, that is, when the effects of
the curvature on the metric of spacetime become significant at the
Planck scale of distances, our arguments probably fail to hold. In all
what follows, we shall therefore always assume that the proper
acceleration $a$ is sufficiently large.

\begin{figure}[htb!]
\begin{center}
\includegraphics{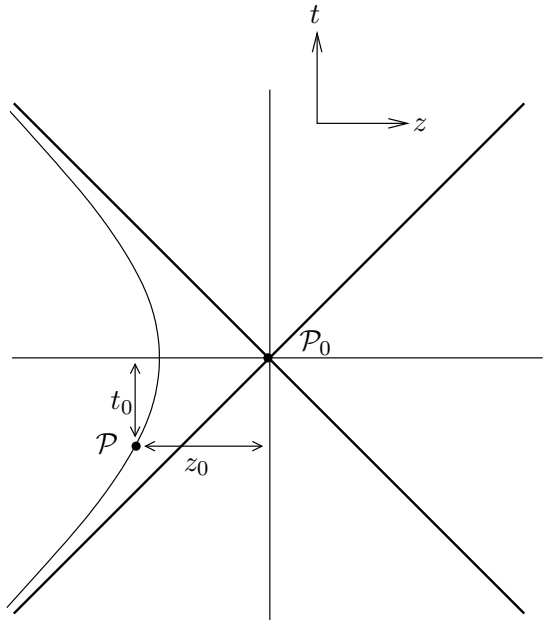}
\caption{Geometrical interpretations of the constants $t_0$ and $z_0$.
  In this figure, the world line of the accelerated two-plane (or,
  equivalently, the world line of the accelerated observer) 
  going through $\mathcal{P}$ is drawn in the frame of
  reference equipped with the geodesic coordinates $t$ and 
  $z$. The origin of the coordinates $t$ and $z$ should lie at the
  point $\mathcal{P}$. The past and the future Rindler horizons of
  the plane are the thick lines which intersect at the point
  $\mathcal{P}_0$. The constant $t_0$ is then the value of the
  coordinate $t$ at the point $\mathcal{P}_0$, whereas the constant
  $z_0$ is the value of the coordinate $z$ at $\mathcal{P}_0$.
  \label{fig:constants2}} 
\end{center}
\end{figure}

The equation for the world line of our plane may be now written as 
\be \label{eq:apaja}\big( z-z_0\big)^2 -\big( t-t_0 \big)^2 = \frac{1}{a^2},
\ee
where $z_0$ and $t_0$ are constants depending on the values of $a$ and
$v$ at the point $\mathcal{P}$. In the (flat) tangent space of the
point $\mathcal{P}$, these constants have solid geometrical
interpretations (see Fig. \ref{fig:constants2}). 
Equation (\ref{eq:apaja}) gives the equation of the world line of the
plane in an immediate vicinity of the point $\mathcal{P}$ with respect
to the orthonormal geodesic coordinates $t$, $x$, $y$, and $z$. If we
solve $z$ from Eq. (\ref{eq:apaja}) and differentiate $z$ with respect
to the time coordinate $t$, we find that the velocity of the plane is,
as a function of the time $t$,
\be \frac{dz}{dt} = \frac{a\, (t_0 -t)}{\sqrt{1+a^2(t_0 -t)^2}}.
\ee
Hence, at the point $\mathcal{P}$, the velocity of the plane is 
\be \label{eq:niitti} v= \frac{at_0}{\sqrt{1+a^2t_0^2}}.
\ee
It is convenient to write the velocity $v$ by means of a new parameter
$\epsilon \in (0,1)$ \cite{Dop} such that
\be \label{eq:nopeus} v = \frac{1-\epsilon}{1+\epsilon},
\ee
and it follows from Eq. (\ref{eq:niitti}) that the constant $t_0$
may be expressed in terms of $\epsilon$ and $a$ as
\be \label{eq:harry} t_0 = \frac{1-\epsilon}{a\sqrt{2\epsilon}}.
\ee
As one may observe, for fixed $a$ the quantity $t_0$ goes to infinity when
$\epsilon$ goes to zero.

Now, what shall be the role of the parameter $\epsilon$ in our analysis? 
We see from Eq. (\ref{eq:nopeus}) that $\epsilon$ describes the
velocity of our plane at $\mathcal{P}$ with respect to the given
system of coordinates. In the limit, where $\epsilon =1$, the plane is
at rest at the point $\mathcal{P}$. On the other hand, when $\epsilon$ takes
its values within the interval $(0,1)$, the plane has initially a
certain velocity relative to the positive $z$-axis such that in the
limit where $\epsilon \rightarrow 0$, the velocity becomes close to
$1$, the speed of light in the natural units. Obviously, for
sufficiently small $\epsilon$, the plane moves with relativistic
speeds with respect to all matter fields, regardless of the properties
of matter at $\mathcal{P}$. Similar results hold also vice versa:
As $\epsilon$ approaches zero, the velocity of the flow of the matter
fields across the plane approaches the speed of light. We have
previously argued that under these circumstances the flow of heat
vastly dominates other forms of energy transfer (the demonstration 
of this claim will be given in Sec. \ref{sec:flow}). Therefore, we
shall henceforth always require that the parameter $\epsilon$ becomes
very small. Only in this limit, we may always interpret the energy flow
through the plane as heat. As we shall soon see, in this limit the
calculations also turn out relatively simple.

\begin{figure}[htb!]
\begin{center}
\includegraphics{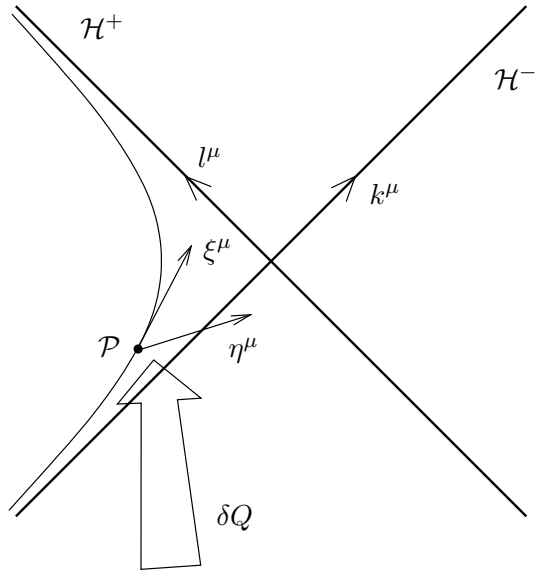}
\caption{The world line of the accelerated spacelike two-plane.
  $\xi^\mu$ is the future pointing unit
  tangent vector of the world line, and $\eta^\mu$ is the spacelike
  unit normal vector of the plane. As one may observe, the world line
  of the plane lies close to its past Rindler horizon
  $\mathcal{H}^-$, which is generated by the null vector $k^\mu$,
  whereas its future Rindler horizon $\mathcal{H}^+$ is generated by
  the null vector $l^\mu$. The large arrow represents the heat that
  flows through the past horizon. 
  \label{fig:heat}}
\end{center}
\end{figure}

So far we have managed to find an appropriate parameter which
determines the velocity of the matter flux across the accelerating
two-plane. It is now time to formulate our ideas by using this
parameter. We denote the future pointing unit tangent vector of the
observer's world line by $\xi^\mu$ and a spacelike unit normal vector
of the plane by $\eta^\mu$. Because the observer, together with the
plane, is assumed to move in the direction perpendicular to the plane,
the vectors $\xi^\mu$ and $\eta^\mu$ are orthogonal. Moreover, we
choose $\eta^\mu$ in such a way that the observer is accelerated in
the direction of the vector $-\eta^\mu$. Since the observer is assumed
to move, at the point $\mathcal{P}$, with the velocity $v$ to the
direction of the positive $z$-axis, the non-zero components of the
vectors $\xi^\mu$ and $\eta^\mu$ are 
{\setlength\arraycolsep{2pt}
\begin{subequations}
\begin{eqnarray} \xi^0 &=& \cosh \phi ,\\ \xi^3 &=& \sinh \phi , \\
  \eta^0 &=& \sinh \phi , \\ \eta^3 &=& \cosh \phi ,
\end{eqnarray}
\end{subequations}}%
where 
\be \phi := \text{arsinh} \Big( \, \frac{v}{\sqrt{1-v^2}} \Big)
\ee
is the boost angle, or rapidity. Using Eq. (\ref{eq:nopeus}) we find:
{\setlength\arraycolsep{2pt}
\begin{subequations} \label{vipu}
\begin{eqnarray} \label{eq:vipu1} \xi^\mu &=& \frac{1}{2} \bigg( \,
  \frac{k^\mu}{\sqrt{\epsilon}} + \sqrt{\epsilon}\, l^\mu  \bigg), \\[6pt]
  \label{eq:vipu2} \eta^\mu &=& \frac{1}{2} \bigg(
  \frac{k^\mu}{\sqrt{\epsilon}} -\sqrt{\epsilon}\, l^\mu  \bigg) ,
\end{eqnarray}
\end{subequations}}%
where $k^\mu := (1,0,0,1)$ and $l^\mu := (1,0,0,-1)$ are null vectors.
This means that when the parameter $\epsilon$ becomes small, the world
line of the observer seems to lie very close to the null geodesic
generated by the null vector $k^\mu$. In the limit, where the proper
acceleration $a$ goes to infinity, the null vector $k^\mu$ becomes a
generator of the past Rindler horizon of the observer moving with the
plane, whereas the null vector $l^\mu$ becomes a generator
of the future Rindler horizon (see Fig. \ref{fig:heat}).

\section{Change of Area}\label{sec:area}
    In the previous Section we considered the trajectory of an observer 
at rest with respect to our accelerating plane in an infinitesimal neighborhood 
of an arbitrary point $\mathcal{P}$ of spacetime. The neighborhood in question was 
assumed be equipped with an orthonormal geodesic system of coordinates. Our main result
was Eq. (\ref{vipu}). That equation told in which way the future directed tangent vector 
$\xi^\mu$ of the observer's world line, as well as the spacelike unit normal vector 
$\eta^\mu$ of the plane may be expressed, at the point $\mathcal{P}$, by means of the
null vectors $k^\mu$ and $l^\mu$, provided that the velocity of the plane is known.
It should be emphasized that Eq. (\ref{vipu}) is always valid, no matter whether spacetime at
the point $\mathcal{P}$ is flat or curved.

    In this Section we proceed to calculate the area change of our plane. In contrast 
to the previous Section, where we considered the world line of an observer at rest 
with respect to the plane we shall, in this Section, consider the world lines of 
\emph{all} the points of the plane. In other words, we shall consider the 
\emph{congruence} of the world lines of the points of our plane. We shall denote
the future directed unit tangent vector field of the smooth congruence under question by 
$\xi^\mu$, and the spacelike unit vector field orthogonal to the plane by $\eta^\mu$, 
because we shall assume that at the point $\mathcal{P}$ the vector fields $\xi^\mu$ and
$\eta^\mu$ will coincide with the vectors $\xi^\mu$ and $\eta^\mu$ of Sec. \ref{sec:tra}. 
At the
point $\mathcal{P}$ we shall use exactly the same geodesic system of coordinates as we 
did in Sec. \ref{sec:tra}. This implies that when our plane lies at the point $\mathcal{P}$, it is
parallel to the $xy$-plane. We shall parametrize the world lines of the points of the 
plane by the proper time $\tau$ measured along these world lines. When the plane lies
at the point $\mathcal{P}$, we have $\tau=0$ for all of the points of the plane. When the 
plane moves away from the point $\mathcal{P}$ and $\tau>0$, the plane is represented by 
the set of points, where $\tau=\text{constant}$ $(>0)$ along the world lines. 

   We shall require that the vector fields $\xi^\mu$ and $\eta^\mu$ have the following 
properties at the point $\mathcal{P}$:
{\setlength\arraycolsep{2pt}
\begin{subequations}\label{ehtoja}
\begin{eqnarray}
\label{eq:ehto1} \xi^\mu &=& \frac{1}{2}\Big( \frac{k^\mu}{\sqrt{\epsilon}} + \sqrt{\epsilon}\,l^\mu \Big),\\
\label{eq:ehto2} \eta^\mu &=&\frac{1}{2}\Big( \frac{k^\mu}{\sqrt{\epsilon}} - \sqrt{\epsilon}\,l^\mu \Big),\\
\label{eq:ehto3} \xi^\mu_{\,\,;1} &=& \xi^\mu_{\,\,;2} = 0,\\
\label{eq:ehto4} a^\mu_{\,\,;1} &=& a^\mu_{\,\,;2} = 0,\\
\label{eq:ehto5} (a^\mu a_\mu)^{1/2} &=& -a^\mu\eta_\mu = a,
\end{eqnarray}
\end{subequations}}%
where 
\begin{equation} \label{eq:amu}
a^\mu := \xi^\alpha\xi^\mu_{\,\,;\alpha}
\end{equation}
is the proper acceleration vector field of our congruence. Equations (\ref{eq:ehto1}) and (\ref{eq:ehto2}),
respectively, are identical to Eqs. (\ref{eq:vipu1}) and (\ref{eq:vipu2}), and they state that our plane 
moves, at the point $\mathcal{P}$, with a certain velocity $v$ determined by the parameter
$\epsilon$ with respect to the geodesic coordinates associated with the point $\mathcal{P}$. 
Eqs. (\ref{eq:ehto3}) and (\ref{eq:ehto4}) involve derivatives with respect
to the coordinates $x$ and $y$ only. Eq. (\ref{eq:ehto3}) states that the tangent
vectors of the world lines of the points of our plane are parallel to each other, whereas 
Eq. (\ref{eq:ehto4}) states that the proper acceleration vectors of the points of the plane are
parallel. Finally, we have Eq. (\ref{eq:ehto5}). That equation states that our plane accelerates, at the point 
$\mathcal{P}$, with a proper acceleration $a$ to the direction opposite to that of the vector $\eta^\mu$.
   
 As a whole our assumptions contained in Eq. (\ref{ehtoja}) correspond to our intuitive picture 
of a material, accelerating plane: When a plane is in an accelerating motion, all its 
points are, initially, moved and accelerated to the same direction, and the possible 
changes in its area result from the curvature of spacetime, rather than from the initial 
conditions posed for the trajectories of its points. It should be emphasized, however, 
that all of the assumptions mentioned in Eq. (\ref{ehtoja}) are just {\it technical} assumptions 
posed for the world lines of the points of our plane. The only really {\it physical} 
assumption of our paper is the hypothesis expressed in Eq. (\ref{eq:rissi}).

   Since our plane is parallel to the $xy$-plane, when $\tau=0$ and the plane lies at 
the point $\mathcal{P}$, the points of the plane have $z=0$, when $\tau=0$. At the point $\mathcal{P}$ 
the change of the area $A$ of our very small plane during an infinitesimal proper time interval $d\tau$ is therefore:
\begin{equation}
dA = A(\xi^1_{\,\,;1} + \xi^2_{\,\,;2})\,d\tau.
\end{equation}
Hence we find, using Eq. (\ref{eq:ehto3}):
\begin{equation} \label{eq:zeroA}
\frac{dA}{d\tau} = 0,
\end{equation}
when $\tau=0$ and the plane lies at the point $\mathcal{P}$. In other words, the first
proper time derivative of the area vanishes, when $\tau=0$. The {\it second} proper
time derivative of the area, however, does not necessarily vanish, when $\tau=0$. We have:
\begin{equation}
\frac{d^2A}{d\tau^2}\Big|_{\tau=0} = A(\dot{\xi}^1_{\,\,;1} + \dot{\xi}^2_{\,\,;2}),
\end{equation}
where the dot means proper time derivative such that $\dot{\xi}^\mu_{\,\,;\nu}:=\frac{d}{d\tau}(\xi^\mu_{\,\,;\nu}) $. 
It is shown in the Appendix that when Eq. (\ref{ehtoja})
holds, we have:
{\setlength\arraycolsep{2pt}
\begin{subequations} \label{tulokset}
\begin{eqnarray}
\label{eq:tulos1} \dot{\xi}^1_{\,\,;1} &=& R^1_{\,\,\mu\nu 1}\xi^\mu\xi^\nu,\\
\label{eq:tulos2} \dot{\xi}^2_{\,\,;2} &=& R^2_{\,\,\mu\nu 2}\xi^\mu\xi^\nu,\\
\label{eq:tulos4} R_{\mu\nu}\xi^\mu\xi^\nu &=& \dot{\xi}^1_{\,\,;1} + \dot{\xi}^2_{\,\,;2} 
+ R^\alpha_{\,\,\mu\nu\beta}\eta_\alpha\eta^\beta\xi^\mu\xi^\nu.
\end{eqnarray}
\end{subequations}}
where $R^\alpha_{\,\,\mu\nu\beta}$ and $R_{\mu\nu}$, respectively, are the Riemann and the
Ricci tensors. So we see that we may write, in general:
\begin{equation} \label{eq:winA1}
\frac{d^2A}{d\tau^2}\Big|_{\tau=0} = AR_{\mu\nu}\xi^\mu\xi^\nu 
- AR^\alpha_{\,\,\mu\nu\beta}\eta_\alpha\eta^\beta\xi^\mu\xi^\nu.
\end{equation}

   Equation (\ref{eq:winA1}) tells in which way the second proper time derivative of the area of our 
accelerating plane depends on the geometry of spacetime. The first special case of 
interest is the one, where the plane is at rest at the point $\mathcal{P}$, and 
spacetime is isotropic in the neighborhood of the point $\mathcal{P}$, i.e. it expands and contracts 
in exactly the same ways in all spatial directions. In that case we have
\begin{equation}
\eta^\mu = \delta^\mu_3,
\end{equation}
which implies that
\begin{equation}
R^\alpha_{\,\,\mu\nu\beta}\eta_\alpha\eta^\beta\xi^\mu\xi^\nu = R^3_{\,\,\mu\nu 3}\xi^\mu\xi^\nu,
\end{equation}
and because spacetime is assumed to be isotropic, we have:
\begin{equation}
R^1_{\,\,\mu\nu 1}\xi^\mu\xi^\nu = R^2_{\,\,\mu\nu 2}\xi^\mu\xi^\nu = R^3_{\,\,\mu\nu 3}\xi^\mu\xi^\nu.
\end{equation}
Hence Eqs. (\ref{tulokset}) and (\ref{eq:winA1}) imply:
\begin{equation} \label{eq:risto}
\frac{d^2A}{d\tau^2}\Big|_{\tau=0} = \frac{2}{3}AR_{\mu\nu}\, \xi^\mu\xi^\nu.
\end{equation}

 Another special case of interest is the one, where the parameter $\epsilon$ gets close 
to zero, and our plane moves with an enormous velocity with respect to the orthonormal
geodesic system of coordinates associated with the point $\mathcal{P}$. Using Eq. (\ref{vipu})
and the symmetry properties of the Riemann and the Ricci tensors we find that we may write
for general $\epsilon>0$:
\begin{equation} \label{eq:winA2}
\frac{d^2A}{d\tau^2}\Big|_{\tau=0} = \frac{1}{4\epsilon}AR_{\mu\nu}k^\mu k^\nu 
+ \frac{1}{2}AR_{\mu\nu}k^\mu l^\nu 
- \frac{1}{4}AR_{\alpha\mu\nu\beta}k^\alpha l^\mu l^\nu k^\beta 
+ \frac{\epsilon}{4}AR_{\mu\nu}l^\mu l^\nu.
\end{equation}
It is easy to see that when $\epsilon\rightarrow 0$, the first term on the right hand 
side of Eq. (\ref{eq:winA2}) will dominate. Hence we may write, for very small $\epsilon$:
\begin{equation} \label{eq:winA3}
\frac{d^2A}{d\tau^2}\Big|_{\tau=0} = \frac{1}{4\epsilon}AR_{\mu\nu}k^\mu k^\nu + \mathcal{O}(1),
\end{equation}
where $\mathcal{O}(1)$ denotes the terms, which are of the order $\epsilon^0$, or higher.
The negative of the right hand side of Eq. (\ref{eq:winA3}) gives the rate of change in the shrinking 
speed of the plane.

\section{Flow of Heat}\label{sec:flow}
    So far we have considered the properties of our accelerating plane only. We shall 
now turn our attention to the matter fields. In general, the boost energy flow of 
matter, or the boost energy flown per unit time through a very small plane with area 
$A$ is, from the point of view of an observer at rest with respect to the plane:
\begin{equation}
\frac{dE_b}{d\tau} = -AT_{\mu\nu}\xi^\mu\eta^\nu,
\end{equation}
where, as in the previous Sections, $\xi^\mu$ is the timelike unit tangent vector of 
the worldline of the observer, and a $\eta^\mu$ is a spacelike unit normal vector of 
the plane, orthogonal to $\xi^\mu$. The negative sign comes from the fact that our plane
is assumed to move, with respect to the matter fields, to the direction of the vector
$\eta^\mu$, and that vector also determines the direction of the boost energy flow 
through our plane.

     When the plane is in an accelerating motion, the boost energy flow through the plane
is not constant, but it will change in time. The rate of change in the boost energy flow 
is:
\begin{equation} \label{eq:Eb}
\frac{d^2E_b}{d\tau^2} = -\dot{A}T_{\mu\nu}\xi^\mu\eta^\nu 
                         - A\dot{T}_{\mu\nu}\xi^\mu\eta^\nu
                         - AT_{\mu\nu}\dot{\xi}^\mu\eta^\nu
                         - AT_{\mu\nu}\xi^\mu\dot{\eta}^\nu,
\end{equation}
where the dot means the proper time derivative. It follows from Eq. (\ref{eq:zeroA}) that when
$\tau=0$, which means that our plane lies at the point $\mathcal{P}$, the first term
on the right hand side of Eq. (\ref{eq:Eb}) vanishes. Because, in general,
{\setlength\arraycolsep{2pt}
\begin{subequations}
\begin{eqnarray}
\dot{\xi}^\mu &=& -a\eta^\mu,\\
\dot{\eta}^\mu &=& -a\xi^\mu
\end{eqnarray}
\end{subequations}}
for an observer moving with a proper acceleration $a$ on the left hand side of the Rindler
wedge, we may write Eq. (\ref{eq:Eb}), by means of the chain rule, as:
\begin{equation} \label{eq:Eb2}
\frac{d^2E_b}{d\tau^2} = \frac{d^2E_{b,t}}{d\tau^2} + \frac{d^2E_{b,a}}{d\tau^2},
\end{equation}
where we have denoted:
\begin{subequations}
\begin{eqnarray}
\frac{d^2E_{b,t}}{d\tau^2} &:=& -AT_{\mu\nu,\alpha}\xi^\mu\eta^\nu\xi^\alpha,\\
\frac{d^2E_{b,a}}{d\tau^2} &:=& aAT_{\mu\nu}(\xi^\mu\xi^\nu + \eta^\mu\eta^\nu).
\end{eqnarray}
\end{subequations}
All quantities have been calculated at the point $\mathcal{P}$.

   The first term on the right hand side of Eq. (\ref{eq:Eb2}) is now due to the simple 
fact that the energy momentum stress tensor $T_{\mu\nu}$ of the matter fields may be different
in different points of spacetime. That term has nothing to do with the acceleration of
the plane. The second term, in turn, is due to the mere acceleration 
of the plane: When the plane is in an accelerating motion, the velocity of the plane with 
respect to the matter fields changes as a function of the proper time $\tau$. In what 
follows, we shall focus our attention to the second term.

   The question is now: In which cases will the second term on the right hand side of 
Eq. (\ref{eq:Eb2}) give the rate of change in the flow of {\it heat}, caused by the mere 
acceleration of the plane? After all, we assumed in our hypothesis that the rate of 
change in the boost energy flow is exactly the rate of change in its heat flow. We shall
see in the next Section that at least in the special case, where the matter consists of 
massless, non-interacting radiation (electromagnetic radiation, for instance) in 
thermal equilibrium, the second term on the right hand side of Eq. (\ref{eq:Eb2}) does indeed
give the rate of change in the heat energy flow through our plane. Another special case is
the one, where there is, instead of massless non-interacting radiation in thermal 
equilibrium, a steady flow of thermal non-interacting massless particles, all propagating
to the one and the same direction. For instance, we may put a source of light to the
focus of a parabolic mirror. In that case the photons reflected from the mirror will all
propagate to the one and the same direction. The photons come out from the 
source of light with all the possible wave lengths, and the energy density
\begin{equation} \label{eq:termo1}
\rho = T_{\mu\nu}\xi^\mu\xi^\nu
\end{equation}
of the photon gas from the point of view of an observer at rest with respect to our accelerating plane
depends on the absolute temperature $T$ of the gas only. If the photons propagate to the 
direction orthogonal to the plane, the pressure exerted by the photons against the plane 
is, in the rest frame of the plane:
\begin{equation}
P = T_{\mu\nu}\eta^\mu\eta^\nu = \rho.
\end{equation}
In other words, the energy density and the pressure of the photon gas under consideration
are equals.

   It is easy to see that the second term on the right hand side of Eq. (\ref{eq:Eb2}) really gives 
the rate of change in the flow of {\it heat} through an accelerating plane for the photon
gas described above. According to the first law of thermodynamics the change in the heat
content of a system with total energy $E$ and pressure $P$ is:
\begin{equation} \label{eq:termo3}
\delta Q = dE + P\,dV,
\end{equation}
where $V$ is the volume of the system. Because the energy density $\rho$ of our photon 
gas is independent of its volume $V$, its total energy is
\begin{equation}
E = \rho V,
\end{equation}
and therefore it follows from Eqs. (\ref{eq:termo1})--(\ref{eq:termo3}) that the heat energy 
possessed by our photon gas per unit volume is
\begin{equation}
\frac{\delta Q}{dV} = T_{\mu\nu}(\xi^\mu\xi^\nu + \eta^\mu\eta^\nu).
\end{equation}
Hence we find that the rate of change in the heat energy flow through our accelerating 
plane is, from the point of view of an observer at rest with respect to our plane:
\begin{equation}
\frac{\delta^2 Q}{d\tau^2} = aAT_{\mu\nu}(\xi^\mu\xi^\nu + \eta^\mu\eta^\nu),
\end{equation}
which is exactly the second term on the right hand side of Eq. (\ref{eq:Eb2}). It should be noted 
that the same result holds whenever the matter consists of massless, non-interacting 
particles only (not just photons), all propagating to the one and the same direction.

  It is remarkable that when our accelerating plane moves with respect to the matter 
fields with a velocity very close to that of light, then the components of the 
energy-momentum stress tensor $T^{\mu\nu}$ of the matter fields behave, from the point
of view of an observer at rest with respect to the plane, in exactly the same way as do
the components of $T^{\mu\nu}$ for matter consisting solely of massless, non-interacting 
particles, all moving in a direction perpendicular to the plane. In other words, in the rest 
frame of a plane moving with an enormous velocity we may consider arbitrary matter, 
in effect, as a gas of non-interacting massless particles.

   To see how this important result comes out, let us fix the rest frame of our plane
at the point $\mathcal{P}$ such that its $z$-axis coincides with the vector $\eta^\mu$,
and its $t$-axis with the vector $\xi^\mu$. In this frame the relevant components of the 
energy momentum stress tensor of arbitrary matter are:
{\setlength\arraycolsep{2pt}
\begin{subequations} \label{teet}
\begin{eqnarray}
T'^{00} &=& T_{\mu\nu}\xi^\mu\xi^\nu,\\
T'^{33} &=& T_{\mu\nu}\eta^\mu\eta^\nu,\\
T'^{03} =T'^{30} &=& -T_{\mu\nu}\xi^\mu\eta^\nu,
\end{eqnarray}
\end{subequations}}%
where $T_{\mu\nu}$ denotes the components of the energy momentum stress tensor in the 
orthonormal geodesic system of coordinates used in Secs. \ref{sec:tra} and \ref{sec:area}. 
For matter consisting solely of massless, non-interacting particles carrying no angular momentum
and propagating to the direction orthogonal to the plane these are the only non-zero
components of $T'^{\mu\nu}$, and we have, in the natural units:
\begin{equation} \label{eq:rho}
T'^{00} = T'^{33} = -T'^{03} = -T'^{30} = \rho,
\end{equation}
where $\rho$ is the energy density of the matter in the rest frame of the plane.

   Now, suppose that instead of having a gas of non-interacting massless particles, the
matter fields are arbitrary. In that case it follows from Eqs. (\ref{vipu}) and (\ref{teet}) that we 
have, in the rest frame of the plane:
{\setlength\arraycolsep{2pt}
\begin{subequations}
\begin{eqnarray}
T'^{00} &=& \frac{1}{4\epsilon}T_{\mu\nu}k^\mu k^\nu + \mathcal{O}(1),\\
T'^{33} &=& \frac{1}{4\epsilon}T_{\mu\nu}k^\mu k^\nu + \mathcal{O}(1),\\
T'^{03} =T'^{30} &=& -\frac{1}{4\epsilon}T_{\mu\nu}k^\mu k^\nu + \mathcal{O}(1),
\end{eqnarray}
\end{subequations}}
where $\mathcal{O}(1)$ denotes the terms, which are of the order $\epsilon^0$, or higher.
So we see that, in the limit, where $\epsilon\rightarrow 0$, which means that the plane
moves with a velocity very close to the speed of light with respect to the matter fields, 
we have:
\begin{equation}
T'^{00} = T'^{33} = -T'^{03} = -T'^{30} = \rho,
\end{equation}
which is exactly Eq. (\ref{eq:rho}). In other words, in the rest frame of a plane moving with an
enormous velocity with respect to the matter fields, the components of the energy momentum
stress tensor are exactly the same as are its components for massless, non-interacting
particles moving in a direction perpendicular to the plane, independently of the kind of 
matter we happen to have. This means that we may consider arbitrary matter, from the point
of view of an observer moving with respect to the matter fields with a very great speed, as
a gas of non-interacting massless particles. Actually, this is something one might expect:
When an observer moves with a very high speed with respect to the matter fields, the 
particles of the matter fields move with respect to the observer with velocities close to 
that of light, and their kinetic energies vastly exceed all the other forms of energy
(mass-energy, for instance). As a result the observer will see, in effect, a gas of 
non-interacting massless particles.

     Our investigations imply that in the limit where the velocity of an accelerating 
plane with respect to the matter fields gets close to the speed of light, the boost energy
flow becomes to the heat flow, independently of the kind of matter we happen to have. In
this limit the second term on the right hand side of Eq. (\ref{eq:Eb2}) gives that part of the rate 
of change in the heat flow which is caused by the mere acceleration of the plane. Using 
Eq. (\ref{vipu}) we find that in the high speed limit the second term on the right hand side of 
Eq. (\ref{eq:Eb2}) may be written as:
\begin{equation} \label{eq:416}
aAT_{\mu\nu}(\xi^\mu\xi^\nu + \eta^\mu\eta^\nu) 
= \frac{1}{2\epsilon}aAT_{\mu\nu}k^\mu k^\nu + \mathcal{O}(\epsilon),
\end{equation}
where $\mathcal{O}(\epsilon)$ denotes the terms, which are of the order $\epsilon$, or 
higher. It is obvious that in the high speed limit, where $\epsilon\rightarrow 0$, the 
terms proportional to $1/\epsilon$ will dominate. When the second term on the right hand
side of Eq. (\ref{eq:Eb2}) gives that part of the rate of change in the heat flow which is caused
by the mere acceleration, we define the rate of change in the {\it acceleration entropy
flow} as:
\begin{equation} \label{eq:417}
\frac{d^2S_a}{d\tau^2}\Big|_{\tau=0} := \frac{1}{T}aAT_{\mu\nu}(\xi^\mu\xi^\nu + \eta^\mu\eta^\nu),
\end{equation}
where $T$ is the absolute temperature, in the rest frame of the plane, of the matter 
flowing through the plane. Using Eq. (\ref{eq:416}) we find that in the high speed limit Eq. (\ref{eq:417})
takes the form:
\begin{equation} \label{eq:418}
\frac{d^2S_a}{d\tau^2}\Big|_{\tau=0} = \frac{1}{T}\frac{1}{2\epsilon}aAT_{\mu\nu}k^\mu k^\nu 
+ \mathcal{O}(\epsilon).
\end{equation}

\section{Massless, Non-Interacting Radiation Fields}\label{sec:rad}
    So far we have managed to find explicit expressions for the both sides of Eq. (\ref{eq:rissi}). 
Its right hand side was obtained in two important cases in Sect. \ref{sec:area}, whereas its
left hand side was obtained in Sect. \ref{sec:flow}. We shall now equate these both sides 
and see,
whether Einstein's field equation really follows from the thermodynamical hypothesis
of Eq. (\ref{eq:rissi}).

    In this Section we shall derive Einstein's field equation in the special case, where 
matter consists solely of massless, non-interacting radiation fields, which are in thermal 
equilibrium in the rest frame of our plane. A typical example of a massless, 
non-interacting radiation field is, of course, the electromagnetic field. Whatever 
massless non-interacting radiation field in thermal equilibrium in the rest frame of our
plane we may have, its energy density is always, in the rest frame of the plane,
\begin{equation} \label{eq:termo51}
\rho = T_{\mu\nu}\xi^\mu\xi^\nu,
\end{equation}
and its pressure 
\begin{equation}
P = T_{\mu\nu}\eta^\mu\eta^\nu
\end{equation}
has the property:
\begin{equation} \label{eq:termo53}
P = \frac{1}{3}\rho.
\end{equation}
It is an important property of the radiation field described above that its energy 
momentum stress tensor $T_{\mu\nu}$ is {\it traceless}, i.e., 
\begin{equation} \label{eq:trace}
T^\alpha_{\,\,\alpha} = 0.
\end{equation}
This property will play an important role in our derivation of Einstein's field equation 
in  the special case considered in this Section.

   The first task is to check, whether the radiation fields under consideration really 
satisfy the assumptions of our hypothesis. In other words, we must check, 
whether the rate of change in the boost energy flow of our radiation is 
really the rate of change in the {\it heat} flow. Obviously, this is
the case: It is a well-known result of elementary thermodynamics that the entropy density
(entropy per unit volume) of massless, non-interacting radiation in thermal equilibrium
is \cite{mandl}
\begin{equation}
s = \frac{1}{T}\frac{4}{3}\rho,
\end{equation}
where $T$ is the absolute temperature of the radiation. So we find that the rate of 
change in the flow of entropy carried by radiation through our plane is
\begin{equation} \label{eq:nakki}
\frac{d^2 S_a}{d\tau^2}\Big|_{\tau=0} = \frac{1}{T}\frac{4}{3}aA\rho.
\end{equation}
It is easy to see that this is exactly the same result as the one given by Eq. (\ref{eq:417}), 
when we use Eqs. (\ref{eq:termo51})--(\ref{eq:termo53}). Hence we may conclude that Eq. (\ref{eq:417}) 
really gives the rate of change in the acceleration entropy flow through our plane for our radiation fields. 
Reasoning backwards and using Eq. (\ref{eq:sinikka1}) then implies that the rate of change in the 
boost energy flow is the rate of change in the heat flow, as required. In other words, 
the assumptions of our hypothesis are satisfied. 
Eqs. (\ref{eq:termo51}) and (\ref{eq:nakki}) imply:
\begin{equation}
\frac{d^2S_a}{d\tau^2}\Big|_{\tau=0} = \frac{1}{T}\frac{4}{3}aAT_{\mu\nu}\xi^\mu\xi^\nu.
\end{equation}
If the absolute temperature $T$ of the radiation agrees with the Unruh temperature $T_U$
of Eq. (\ref{eq:unruhT}), our final expression for the acceleration entropy flow takes the form:
\begin{equation} \label{eq:wille}
\frac{d^2S_a}{d\tau^2}\Big|_{\tau=0} = \frac{8\pi}{3}AT_{\mu\nu}\xi^\mu\xi^\nu.
\end{equation}
It should be emphasized, however, that although we have taken the temperature of the 
radiation to agree with the Unruh temperature measured by an observer at rest with respect
to our accelerating plane, Eq. (\ref{eq:wille}) does {\it not} give the acceleration entropy flow of 
the Unruh radiation. Our radiation field is just an ordinary, massless, non-interacting 
radiation field, which has no connection whatsoever with the Unruh radiation. We may either heat up 
or cool down the radiation until its temperature agrees with the Unruh temperature. Our 
claim is that after the Unruh temperature has been reached, the hypothesis of Eq. (\ref{eq:rissi})
holds, and it implies Einstein's field equation.

    After finding an expression for the acceleration entropy flow we shall now turn our 
attention to the rate of change in the shrinking speed of the plane. Since the radiation
is assumed to be in thermal equilibrium with respect to the plane, when the plane lies at 
the point $\mathcal{P}$, it is isotropic in the neighborhood of the point $\mathcal{P}$, 
and spacetime expands and contracts in exactly the same ways in all spatial directions at the 
point $\mathcal{P}$. This means that $R^1_{\,\,\mu\nu 1}\xi^\mu\xi^\nu$, 
$R^2_{\,\,\mu\nu 2}\xi^\mu\xi^\nu$ and $R^3_{\,\,\mu\nu 3}\xi^\mu\xi^\nu$ are equals, 
and we may use Eq. (\ref{eq:risto}). Using Eq. (\ref{eq:wille}) 
for the left hand side, and the negative of the right hand side of Eq.(3.11) for the right hand 
side of Eq. (\ref{eq:rissi}), we find that Eq. (\ref{eq:rissi}) 
implies:
\begin{equation}
\frac{8\pi}{3}AT_{\mu\nu}\xi^\mu\xi^\nu = -\frac{1}{3}AR_{\mu\nu}\xi^\mu\xi^\nu,
\end{equation}
or:
\begin{equation}
R_{\mu\nu}\xi^\mu\xi^\nu = -8\pi T_{\mu\nu}\xi^\mu\xi^\nu.
\end{equation}
Since $\xi^\mu$ is an arbitrary timelike unit vector field, we must have:
\begin{equation}
R_{\mu\nu} = -8\pi T_{\mu\nu},
\end{equation}
which is exactly Einstein's field equation
\begin{equation}
R_{\mu\nu} = -8\pi(T_{\mu\nu} - \frac{1}{2}g_{\mu\nu}T^\alpha_{\,\,\alpha}),
\end{equation}
or
\begin{equation} \label{eq:tracelessEFE}
R_{\mu\nu} - \frac{1}{2}g_{\mu\nu}R = -8\pi T_{\mu\nu}
\end{equation}
in the special case, where the tensor $T_{\mu\nu}$ is {\it traceless}, i.e., Eq. (\ref{eq:trace}) 
holds. Since $T_{\mu\nu}$ is indeed traceless for massless, non-interacting radiation 
fields in thermal equilibrium, we have managed to obtain Einstein's field equation from
our hypothesis for such radiation fields.

\section{General Matter Fields}\label{sec:EFE}
   We saw in the previous Section how Einstein's field equation follows from our hypothesis 
(\ref{eq:rissi}) concerning the thermodynamical properties of spacetime and matter fields at least
in the special case, where matter consists of massless, non-interacting radiation in thermal 
equilibrium only. An advantage of such radiation is that the rate of change in its boost energy flow through our
accelerating plane is exactly the rate of change in the heat flow, and therefore the assumptions of our 
hypothesis are automatically satisfied. When attempting to generalize the thermodynamical
derivation of Einstein's field equation of the Sec. \ref{sec:rad} for general matter fields, 
however, one meets with difficulties, because for general matter fields the boost energy
flow may include other forms of energy, except heat as well (mass energy, for instance), 
and hence the assumptions of our hypothesis are not necessarily satisfied for general
matter fields.

   Fortunately, we managed to show in Sec. \ref{sec:flow} that there is a way out of this problem:
We take our accelerating plane to move, at the point $\mathcal{P}$ under consideration, 
with a velocity $v$ very close to the speed of light with respect to the matter fields. 
From the point of view of an observer moving with an enormous velocity with respect to the
matter fields all matter behaves, in effect, as a gas of non-interacting massless 
particles, and the rate of change in the boost energy flow equals with the rate of 
change in the heat flow. Hence we may apply our 
hypothesis in the limit, where $v$ gets close to 1, the speed of light in the natural
units. In this limit the rate of change in the shrinking speed of the plane is given
by Eq. (\ref{eq:winA3}) and the rate of change in the acceleration entropy flow
is given by Eq. (\ref{eq:418}). Using Eq. (\ref{eq:winA3}) for the left hand side, and 
the negative of the right hand side of Eq. (\ref{eq:rho}) for the 
right hand side of Eq. (\ref{eq:rissi}) we find:
\begin{equation}
\frac{1}{T}\frac{a}{2\epsilon}AT_{\mu\nu}k^\mu k^\nu + \mathcal{O}(\epsilon)
= -\frac{1}{8\epsilon}AR_{\mu\nu}k^\mu k^\nu + \mathcal{O}(1).
\end{equation}
Again, if the absolute temperature $T$ of the matter agrees with the Unruh temperature 
$T_U$ of Eq. (\ref{eq:unruhT}) we get, in the high speed limit, where $\epsilon\rightarrow 0$:
\begin{equation}
R_{\mu\nu}k^\mu k^\nu = -8\pi T_{\mu\nu}k^\mu k^\nu.
\end{equation}
Since $k^\mu$ is an arbitrary, future directed null vector field, we have:
\begin{equation}
R_{\mu\nu} + fg_{\mu\nu} = -8\pi T_{\mu\nu},
\end{equation}
where $f$ is some function of the spacetime coordinates. It follows from the Bianchi 
identity
\begin{equation}
(R^\mu_{\,\,\nu} - \frac{1}{2}R\delta^\mu_\nu)_{;\mu} = 0,
\end{equation}
that $f=-\frac{1}{2}R + \Lambda$ for some constant $\Lambda$, and hence we arrive at the
equation
\begin{equation} \label{eq:totEFE} R_{\mu\nu} - \frac{1}{2}g_{\mu\nu}R + \Lambda g_{\mu\nu} = -8\pi T_{\mu\nu},
\end{equation}
which is Einstein's field equation with the cosmological constant $\Lambda$.
 
\section{Concluding Remarks} \label{sec:koonti}
In this paper we have obtained Einstein's field equation 
by means of very simple thermodynamical arguments 
concerning the properties of a very small spacelike two-plane in a uniformly 
accelerating motion. Our derivation was based on a hypothesis that
when matter flows through the plane, 
and the temperature of the matter is the same as the Unruh
temperature measured by an observer at rest with respect to the plane, then the rate 
of change in the flow of entropy caused by the mere acceleration of the plane
is, in natural units, exactly one-half of the rate of change in the shrinking
speed of the area of the plane. From 
this hypothesis we obtained, by means of the fundamental thermodynamical relation
$\delta Q = T\,dS$, Einstein's field equation.

When spacetime is filled with isotropic, massless, non-self-interacting
radiation field (electromagnetic field, 
for instance) in thermal equilibrium, it is very easy to obtain Einstein's field equation
from our hypothesis, because it turns out that in this case the rate of change in the
boost energy flow through the plane is exactly the rate of change in the heat flow of the 
radiation. However, if 
the fields are massive, or self-interacting, the situation becomes more complicated, 
because the boost energy flow involves other forms of energy, except heat, as well 
(mass-energy, for instance). In that 
case we may consider the situation, where the plane moves with 
respect to the matter fields with a velocity very close to that of light. When the plane
moves with respect to the matter fields with an enormous velocity, it
turns out that the amount of heat vastly exceeds the amounts of 
other forms of energy carried by matter through the plane, and Einstein's field equation
for general matter fields follows from our hypothesis.

Our derivation of Einstein's field equation by means of purely thermodynamical arguments 
provides support for the idea, earlier expressed by Jacobson, that Einstein's field 
equation may actually be understood as a thermodynamical equation of state of spacetime and 
matter fields. 
Although our thermodynamical derivation of Einstein's field equation bears a lot of 
similarities with Jacobson's derivation, it should be strongly emphasized the radical 
difference between these two derivations: Jacobson considered the boost energy flow through
a \emph{horizon} of spacetime, whereas we considered the boost energy flow through an 
accelerating, spacelike two-plane. Horizons of spacetime are certain null hypersurfaces of
spacetime, and therefore they are created, when all points of a spacelike two-surface move
along certain \emph{null} curves of spacetime. In contrast, our spacelike two-plane 
was assumed to move in spacetime with a speed less than that of light, and therefore all of
its points move along \emph{timelike} curves of spacetime. Because of that our two-plane
should not be considered as a part of any horizon of spacetime. Nevertheless, we found that
if the entropy carried by matter through the plane is connected with the change in its area
in a certain manner, then Einstein's field equation follows.
The fact that an assumption
of a simple proportionality between the rates of changes in the entropy flow and in the 
shrinking speed of the plane yields Einstein's field equation even when that two-plane is 
not a part of any horizon of spacetime strongly suggests that one may associate 
meaningfully the concept of gravitational entropy not only with horizons, but also 
with arbitrary spacelike two-surfaces of spacetime \cite{Fursaev}. It is still uncertain what the 
consequences of such a possibility may be, but they will most likely have some 
influence on our views of the nature of gravitational entropy.           

\begin{acknowledgments} We thank Jorma Louko for his constructive
  criticism during the preparation of this paper.
\end{acknowledgments}

\appendix*

\section{Proof of Eq. (\ref{tulokset})}

     In this Appendix we shall show that Eq. (\ref{ehtoja}) implies Eq. (\ref{tulokset}). It follows from the 
chain rule that, in geodesic coordinates,
\begin{equation}
\dot{\xi}^\mu_{\,\,;\nu} :=\frac{d}{d\tau}(\xi^\mu_{\,\,;\nu}) = \xi^\alpha\xi^\mu_{\,\,;\nu;\alpha}.
\end{equation}
Using the trivial identity
\begin{equation}
\xi^\alpha\xi^\mu_{\,\,;\nu;\alpha} = \xi^\alpha\xi^\mu_{\,\,;\alpha;\nu} 
- \xi^\alpha(\xi^\mu_{\,\,;\alpha;\nu} - \xi^\mu_{\,\,;\nu;\alpha}),
\end{equation}
the product rule of covariant differentiation, and the basic properties of the Riemann 
tensor we get:
\begin{equation}
\dot{\xi}^\mu_{\,\,;\nu} = (\xi^\alpha\xi^\mu_{\,\,;\alpha})_{;\nu} 
- \xi^\alpha_{\,\,;\nu}\xi^\mu_{\,\,;\alpha} - \xi^\alpha R^\mu_{\,\,\alpha\nu\beta}\xi^\beta.
\end{equation}
Eq. (\ref{eq:amu}) and the symmetry properties of the Riemann tensor imply:
\begin{equation}
\dot{\xi}^\mu_{\,\,;\nu} = a^\mu_{\,\,;\nu} - \xi^\alpha_{\,\,;\nu}\xi^\mu_{\,\,;\alpha} 
+ R^\mu_{\,\,\alpha\beta\nu}\xi^\alpha\xi^\beta.
\end{equation}
So we have:
\begin{subequations}
\begin{eqnarray}
\label{eq:A51} \dot{\xi}^1_{\,\,;1} &=& a^1_{\,\,;1} - \xi^\alpha_{\,\,;1}\xi^1_{\,\,;\alpha} 
+ R^1_{\,\,\mu\nu 1}\xi^\mu\xi^\nu,\\
\label{eq:A52} \dot{\xi}^2_{\,\,;2} &=& a^2_{\,\,;2} - \xi^\alpha_{\,\,;2}\xi^2_{\,\,;\alpha} 
+ R^2_{\,\,\mu\nu 2}\xi^\mu\xi^\nu.
\end{eqnarray}
\end{subequations}
Eqs. (\ref{eq:ehto3}) and (\ref{eq:ehto4}) imply that the first two terms on the right hand sides of 
Eqs.(\ref{eq:A51})--(\ref{eq:A52}) will vanish. So we get:
\begin{subequations}
\begin{eqnarray}
\label{eq:A61} \dot{\xi}^1_{\,\,;1} &=& R^1_{\,\,\mu\nu 1}\xi^\mu\xi^\nu,\\
\label{eq:A62} \dot{\xi}^2_{\,\,;2} &=& R^2_{\,\,\mu\nu 2}\xi^\mu\xi^\nu.
\end{eqnarray}
\end{subequations}
which are Eqs. (\ref{eq:tulos1})--(\ref{eq:tulos2}).

  It only remains to prove Eq. (\ref{eq:tulos4}). To this end we note first that the vectors 
$\xi^\mu$, $\eta^\mu$ $e^\mu_{(1)}$, and $e^\mu_{(2)}$, where $e^\mu_{(1)}$ and $e^\mu_{(2)}$, 
respectively, are the spacelike unit vectors parallel to the $x$- and the $y$-axes of
the orthonormal geodesic system of coordinates at the point $\mathcal{P}$, constitute 
an orthonormal set of vectors. In other words, the vectors $\xi^\mu$, $\eta^\mu$, 
$e^\mu_{(1)}$ and 
$e^\mu_{(2)}$ may be taken to be the base vectors of a new orthonormal geodesic system 
of coordinates at the point $\mathcal{P}$. In this system of coordinates our plane is at
rest at the point $\mathcal{P}$, and it has been obtained from the original orthonormal 
geodesic system of coordinates by means of the Lorentz boost. Hence we find, 
by means of the antisymmetry properties
\begin{equation}
R_{\alpha\beta\mu\nu} = -R_{\beta\alpha\mu\nu} = -R_{\alpha\beta\nu\mu}
\end{equation}
of the Riemann tensor that
\begin{equation}
R^1_{\,\,\mu\nu 1}\xi^\mu\xi^\nu + R^2_{\,\,\mu\nu 2}\xi^\mu\xi^\nu 
+ R_{\alpha\mu\nu\beta}\eta^\alpha\eta^\beta\xi^\mu\xi^\nu = R_{\mu\nu}\xi^\mu\xi^\nu,
\end{equation}
where $R^\sigma_{\,\,\mu\nu\sigma}$ is the Ricci tensor. Using Eqs. (\ref{eq:A61})--(\ref{eq:A62}) we 
therefore get:
\begin{equation}
R_{\mu\nu}\xi^\mu\xi^\nu = \dot{\xi}^1_{\,\,;1} + \dot{\xi}^2_{\,\,;2} 
+ R^\alpha_{\,\,\mu\nu\beta}\eta_\alpha\eta^\beta\xi^\mu\xi^\nu,
\end{equation}
which is Eq. (\ref{eq:tulos4}).


\begin{thebibliography}{20}
\bibitem{bek} J.~D.~Bekenstein, Phys. Rev. D \textbf{7}, 2333 (1973).
\bibitem{haw} S.~W.~Hawking, Commun. Math. Phys. \textbf{43}, 199 (1975).
\bibitem{unruh} W.~G.~Unruh, Phys. Rev. D \textbf{14}, 870 (1976).
\bibitem{bath1} G.~W.~Gibbons and S.~W.~Hawking, Phys. Rev. D
  \textbf{15}, 2738 (1977). See also, S.~W.~Hawking, Phys Rev. D \textbf{18}, 1747 (1978).
\bibitem{pad} T.~Padmanabhan, Gen. Rel. Grav. \textbf{34}, 2029
  (2002).  See also T.~Padmanabhan, Class. Quant. Grav. \textbf{21},
  4485 (2004), Phys. Rept. \textbf{406}, 49 (2005), 
and Braz. J. Phys. {\bf 35}, 362 (2005),
  T.~Padmanabhan and A.~Patel arXiv:gr-qc/0309053, D.~Kothawala, S.~Sarkar
  and T.~Padmanabhan, arXiv:gr-qc/0701002, and T.~Padmanabhan and
  A.~Paranjape, Phys. Rev. D {\bf 75}, 064004 (2007). 
\bibitem{jac} T.~Jacobson, Phys. Rev. Lett. \textbf{75}, 1260
  (1995). See also, T.~Jacobson, Found. Phys. \textbf{33}, 323 (2003)
  and T.~Jacobson in \emph{General relativity and Relativistic
    Astrophysics: Eighth Canadian Conference} (AIP Conference
  Proceedings 493, edited by C.~P.~Burgess and R.~C.~Myers (AIP, 1999)).
\bibitem{edd} Curiously, the idea that gravitation may be
  only a statistical effect was proposed already by A.~Eddington in
  \emph{Space, Time and Gravitation} (Syndics of the Cambridge
  University Press, London, 1920). 
\bibitem{tiw} Jacobson's analysis has been recently reconsidered by
  S.~C.~Tiwari, arXiv:gr-qc/0612099. See also, S.~C.~Tiwari, arXiv:0705.3882.
\bibitem{mp1} J.~M\"akel\"a and A.~Peltola, Phys. Rev. D \textbf{69}
  124008 (2004).
\bibitem{mp2} J.~M\"akel\"a and A.~Peltola, arXiv:gr-qc/0406032.
\bibitem{jamo} J.~M\"akel\"a, arXiv:gr-qc/0605098. See also, J.~M\"akel\"a, arXiv:gr-qc/0506087.
\bibitem{thie} Somewhat similar problems also arise in the loop quantum
  gravity, when one tries to define the concept of area for
  ``one-sided'' two-surfaces (i.e. for horizons). This problem has
  been discussed, for instance, by T.~Thiemann in
  \emph{Modern Canonical Quantum General Relativity} (Cambridge
  University Press, Cambridge, 2004) and C.~Rovelli in
  \emph{Quantum Gravity} (Cambridge University Press, Cambridge, 2004).
\bibitem{Dop} Actually, the parameter $\epsilon$ is the square of the
  so-called Doppler shift factor.
\bibitem{mandl} See, for instance, F.~Mandl, \emph{Statistical
    Physics}, 2nd edn. (John Wiley \& Sons Ltd., Chichester, 1988).
\bibitem{Fursaev} Somewhat related ideas have been expressed by D. Fursaev, 
arXiv:0711.1221
\end{thebibliography}
\end{document}